\documentstyle[11pt,aaspp4]{article}

\newcommand\beq{\begin{equation}}
\newcommand\eeq{\end{equation}}

\lefthead{MENOU, PERNA \& RAYMOND}
\righthead{X-RAY LINES FROM HOT FLOWS AROUND WHITE DWARFS}

\def\gsim{\;\rlap{\lower 2.5pt
 \hbox{$\sim$}}\raise 1.5pt\hbox{$>$}\;}
\def\lsim{\;\rlap{\lower 2.5pt
   \hbox{$\sim$}}\raise 1.5pt\hbox{$<$}\;}

\begin{document}

\title{X-Ray Lines from Hot Flows around White Dwarfs. Application to
SS Cygni.}

\author{Kristen Menou,\altaffilmark{1}}

\affil{Princeton University, Department of Astrophysical Sciences,
Princeton NJ 08544, USA, kristen@astro.princeton.edu}

\author{Rosalba Perna\altaffilmark{2} and John C. Raymond}

\affil{Harvard-Smithsonian Center for Astrophysics, 60 Garden Street,
Cambridge MA 02138, USA, rperna@cfa.harvard.edu,
jraymond@cfa.harvard.edu}

\altaffiltext{1}{Chandra Fellow}
\altaffiltext{2}{Harvard Junior Fellow}

\begin{abstract}
Rather than accreting via a disk, some White Dwarfs (WDs) in quiescent
Dwarf Novae (DN) could accrete via an Advection-Dominated Accretion
Flow (ADAF) possibly responsible for the X-ray Bremsstrahlung emission
observed.  Such a hot accretion flow is also expected to produce
characteristic thermal line emission. Using SS Cyg as a specific
example, we show that knowing {\it a priori} the inclination and the
WD mass in quiescent DN makes X-ray line diagnostics powerful probes
of the flow structure in these systems. Current X-ray instruments can
discriminate, from their width, between lines emitted from a flow with
a Keplerian rotation rate and those emitted at a substantially
sub-Keplerian rate. This could be used to observationally test the
property of energy advection, which is at the origin of the partial
radial pressure support by the hot gas and the sub-Keplerian rotation
rate in an ADAF.
\end{abstract}


\section{Introduction}          

Dwarf novae (DN) are members of the class of Cataclysmic Variables
(CVs), which are binary stars composed of a main-sequence donor which
transfers mass via Roche-lobe overflow onto a White Dwarf (WD). DN
experience quasi-regular, luminous outbursts during which accretion onto the
WD proceeds at a high rate. Most of the time, however, DN are in
quiescence, a phase during which the accretion rate onto the WD is
much reduced (see Warner 1995 for a review).

Quiescent DN are well known sources of hard X-ray emission (C\'ordova
\& Mason 1983; Patterson \& Raymond 1985a). This emission, typically
at a level of $\sim 10^{30-32}$~erg~s$^{-1}$, is consistent with
Bremsstrahlung emission from a hot gas with temperatures $\sim
2-20$~keV (Patterson \& Raymond 1985a; Eracleous, Halpern \& Patterson
1991; Belloni et al. 1991; Yoshida, Inoue \& Osaki 1992; Mukai \&
Shiokawa 1993).  It has been commonly attributed to the boundary layer
(BL) of the accreting WD, because at low enough accretion rates
($\lsim 10^{16}$~g~s$^{-1}$), the BL is unable to cool before it
expands out of the plane of the accretion disk, so that it is hot,
optically-thin and therefore a significant source of hard X-rays
(Pringle \& Savonije 1979; Tylenda 1981; King \& Shaviv 1984;
Patterson \& Raymond 1985a).

Detailed calculations by Narayan \& Popham (1993) show that the
optically-thin BLs of accreting WDs are also radially extended (on the
order of the WD radius in their models) and that they advect part of
the internally dissipated energy inwards to the WD surface as a
consequence of their inability to cool. The similarities between the
optically-thin BL solutions of Narayan \& Popham (1993) and ADAF
solutions (Ichimaru 1977; Rees et al. 1982; Narayan \& Yi 1994; 1995;
Abramowicz et al. 1995; see Narayan, Mahadevan \& Quataert 1998 for a
review) prompted Menou (2000) to propose that ADAFs could actually be
present around WDs accreting at low rates, such as in some
quiescent DN.

ADAF models applied to accreting WDs are able to reproduce the typical
X-ray luminosities of quiescent DN, as well as the range of plasma
temperatures inferred from early X-ray observations. In addition,
advection in the ADAF results in a large amount of energy deposited
onto the central WD. If this energy is thermalized in the WD
atmosphere before being radiated away, a strong EUV emission component
originates from the heated atmosphere which can explain the luminous
He~II $\lambda 4686$ lines often observed in quiescent dwarf novae as
being the result of disk irradiation (Menou 2000). In fact, the
presence of such a component had been suggested by previous studies
which emphasized the large supply of EUV photons (difficult to observe
directly) required to explain the strength of the lines by
photoionization of the disk material (Patterson \& Raymond 1985b;
Vrtilek et al. 1994).  Simple BL models (in which one assumes the
kinetic energy of the accreting gas just above the WD surface is
radiated as isothermal, optically-thin Bremsstrahlung emission; e.g.,
Tylenda 1981; Patterson \& Raymond 1985a) do not offer any satisfying
explanation to the strength of these lines (see also, e.g., Ferland et
al. (1982) for a discussion of additional difficulties faced by simple
models).

The success of ADAF models applied to quiescent DN in providing an
explanation for both the hard X-ray emission and the strong He~II
lines is attractive, but not compelling. In particular, X-ray eclipse
observations indicate that a radially extended ADAF is not present in
all quiescent DN (Wood et al. 1995; Mukai et al. 1997; Pratt et
al. 1999; Menou 2000).  In this paper, we emphasize that the presence
of an ADAF can be tested in non-eclipsing systems by detecting (or
not) thermal emission lines from the hot plasma, which are robust
predictions of the theoretical models (Narayan \& Raymond 1999). Line
diagnostics can put severe constraints on the structure of the hot
flow (Perna, Raymond \& Narayan 2000); this is especially true for
quiescent DN, which are systems with a value of the WD mass and the
inclination to the line of sight often relatively well constrained.

Other variations on the theme of a thin Shakura-Sunyaev (1973) disk
with a hot, optically thin boundary layer include the model of King \&
Shaviv (1984), who emphasized the role of thermal conduction and the
qualitative change expected at higher accretion rates. The optically
thick boundary layers of high accretion rate systems produce less
luminosity than expected from the outer disk luminosity (e.g., Mauche,
Raymond \& Mattei 1995; Popham \& Narayan 1995) through some
combination of advection and loss in a wind.

In \S2, we recall some important properties of the DN SS Cyg to be
used as constraints for the models. In \S3, we describe the ADAF
models and the numerical code used for the spectral predictions. In
\S4, we show predictions for the system SS Cyg and we emphasize the
important constraints that can be put on the structure of the hot flow
from detailed X-ray line diagnostics. Finally, we discuss some
consequences and limitations of this work in \S5, before summarizing
our main results in \S6.

\section{SS Cygni and its quiescent X-ray emission}

Observations show that some quiescent DN are more likely to harbor an
ADAF than others.  In particular, of the five DN observed in
quiescence by the Hopkins Ultraviolet Telescope (Long 1996; spectral
range $\sim 830-1860 \AA $), three (WX Hyi, Yz Cnc and SS Cyg) show
strong, blue continua with no evidence for the presence of a normal WD
atmosphere (broad instead of narrow emission lines and no evidence for
the expected broad Ly$\alpha$ absorption feature).  These three
sources are ideal candidates for the presence of an ADAF because a
blue UV continuum is expected if the WD atmosphere is heated up to
several $10,000$~K by a substantial amount of energy advection in the
surrounding hot flow (Menou 2000). SS Cyg is one of the best studied
DN (see, e.g., Warner 1995), a bright source of X-ray emission in
quiescence and is an approved target for observation with the {\it
Chandra X-ray Observatory}. This motivates us to construct specific
models for this system in what follows. The only other X-ray spectral
models of SS Cyg in quiescence that we know of are the one-temperature
models of Done \& Osborne (1997), which are based on the BL solutions
of Narayan \& Popham (1993).

The mass of the WD in SS Cyg is $M_{\rm WD} \simeq 1.2 M_\odot$ and
the inclination of the system to the line of sight is $i \simeq 40^o$
(Ritter \& Kolb 1998). This inclination guarantees that the broadening
of the X-ray lines from an ADAF is dominated by the bulk motion of the
flow rather than thermal and turbulent motions (see below).  The X-ray
emission of SS Cyg during quiescence is variable. It has been
characterized by an X-ray luminosity in the range
$10^{31}-10^{32}$~erg~s$^{-1}$ (in various X-ray bands) and gas
temperatures in the range $5-19$~keV (see, e.g., Yoshida et al. 1992;
Mukai \& Shiokawa 1993).

\section{Modeling techniques}

There are a number of important assumptions made in the models
presented below. For clarity, we briefly recall them here, while a
more detailed discussion of these assumptions and their validity can
be found in Menou (2000).

The hot flow structure used in the present study for the spectral
calculations is nearly identical to the self-similar ADAF solutions
described by Narayan \& Yi (1994). This flow must make a transition
from a large (but still subsonic) radial speed to a zero radial speed
at the stellar surface. It is assumed that this transition happens in
a narrow region (boundary layer, or BL) in the vicinity of the WD
surface. The structure of this BL is not described in any detail
here. The amount of mechanical energy released in this region (which
depends on the stellar rotation rate) is comparable to the amount of
energy radiated by the ADAF (Menou 2000). In all the models presented,
we assume that both this mechanical energy and the much larger amount
of energy advected by the ADAF (see below) are released at the WD
surface as extreme-UV radiation,\footnote{The large amount of energy
advection results in significant heating of the WD and could have
important consequences on the structure of its envelope (e.g. Shaviv
\& Starrfield 1987; Pringle 1988; Regev \& Shara 1989)} so that these
additional components are not considered in the X-ray emission models
presented (broadband models show that the EUV radiation goes to
infinity without interacting significantly with the flow, though
prominent HeII emission lines may be a signature of this component;
Menou 2000). We expect a roughly blackbody spectrum with the
luminosity of the advected energy from the WD surface, but detailed
models of the fraction of the WD surface covered and of the structure
of the heated atmosphere are needed to make more specific predictions.

\subsection{ADAF models}

The radial profiles of physical quantities in the hot flow, such as
density, radial and azimuthal velocities and temperature are obtained
from ADAF models and then used as an input for the detailed spectral
calculations described in \S3.2. We use two specific ADAF models
presented by Menou (2000), which have parameters appropriate for SS
Cyg: $M_{\rm WD}=1.2 M_\odot$ (WD mass), $R_{\rm WD} =5 \times
10^8$~cm (WD radius), $\alpha_{\rm ADAF}=0.2$ (viscosity parameter),
$\beta=6$ (ratio of gas to magnetic pressure), $\gamma=1.636$
(adiabatic index of the fluid, which includes contributions from the
particles, the turbulence and the magnetic field; see Quataert \&
Narayan 1999), $\delta=10^{-2}$ (fraction of direct electron viscous
heating; unimportant for the present models), $p=0$ (no wind) and
$\dot M = 7 \times 10^{-3} \dot M_{\rm Edd}$ (accretion rate), where
$\dot M_{\rm Edd} = 1.39 \times 10^{18}~(M_{\rm
WD}/M_\odot)$~g~s$^{-1}$ is the Eddington accretion rate for a $10\%$
radiative efficiency. In the two models, this value of $\dot m$
corresponds to a physical accretion rate $\dot M \simeq
10^{16}$~g~s$^{-1}$~$\simeq 1.5 \times 10^{-10}~M_\odot$~yr$^{-1}$, a
$0.5-10$~keV X-ray luminosity of $\simeq 2 \times
10^{31}$~erg~s$^{-1}$ and $\approx 85$\% of the total gravitational
energy liberated being advected by the flow.

The difference between the two models comes from the radial extent of
the ADAF. In model I, the ADAF extends from the WD surface at $R_{\rm
WD}$ up to $R_{\rm max}=10^4 R_s$, where $R_s= 3 \times 10^5 (M_{\rm
WD}/M_\odot)$~cm is the Schwarzschild radius. In model II, the ADAF
extends further out, up to $R_{\rm max}=10^5 R_s$. In addition, the
accretion rate in the ADAF decreases as $R_{\rm trans}/R$ beyond
$R_{\rm trans}=10^4 R_s$ in model II, to represent a corona accreting
gas that is gradually being evaporated from an underlying thin disk
(see Menou 2000 and Esin, McClintock \& Narayan 1997 for details). In
the following, we neglect the role, even indirect, that the disk could
have on the X-ray spectrum emitted by the system (see \S5 for a
discussion of a possible reflection component).  Except for the
spectral calculations, all the other model characteristics are similar
to those described in Menou (2000).

\begin{figure}
\plotone{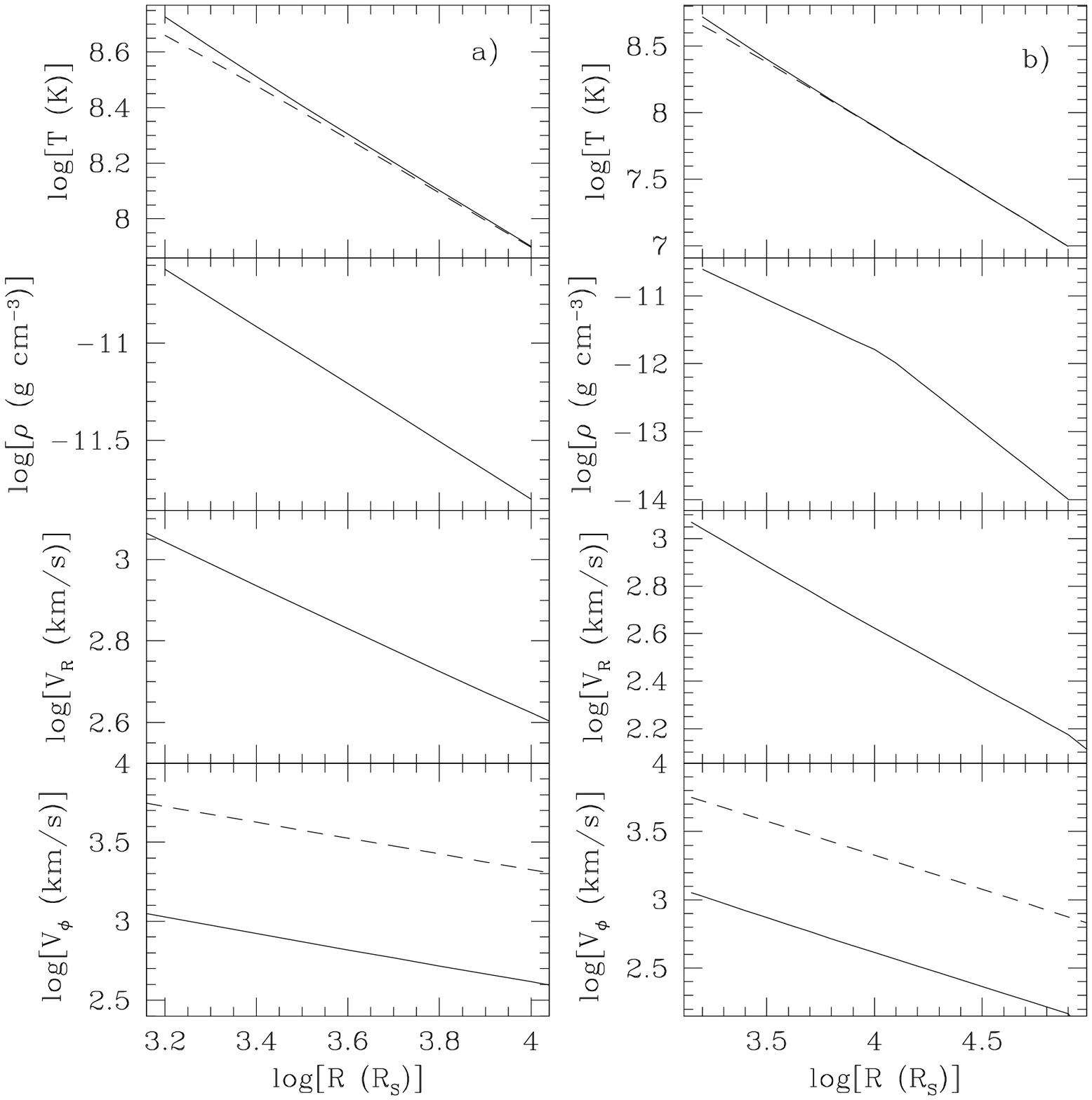}
\caption{The radial profiles in ADAF models I (a) and II (b) of
quantities that are important for the spectral calculations. From top
to bottom, the panels show the gas temperature (solid: ions, dashed:
electrons), the gas density, the radial infall speed and the azimuthal
speed (solid: value for the ADAF, dashed: Keplerian value).  Profiles
are shown as a function of radius in Schwarzschild units, from
$10^{3.15}$ ($=R_{\rm WD}=5 \times 10^8$~cm) to $10^4$ ($\approx 7
R_{\rm WD}$) or $10^5$ ($\approx 70 R_{\rm WD}$)
\label{fig:one}}
\end{figure}

Panels (a) and (b) of Figure 1 show the radial profiles of various
quantities in models~I and~II, respectively. At radii $\ga 10^3 R_s$,
the ions and the electrons in the ADAF have essentially the same
temperature (because of an efficient coupling by Coulomb collisions),
from a few $10^8$~K close to the WD to $\sim 10^7$~K further
away. This range of temperatures is ideal for the efficient production
of thermal X-ray emission lines (Narayan \& Raymond 1999). The radial
profiles closely follow the self-similar ADAF solutions of Narayan \&
Yi (1994).  A change in the slope of the density profile is clearly
visible in Figure~\ref{fig:one}b; it corresponds to the radius $R_{\rm
trans}$ beyond which the mass accretion rate in the ADAF starts
decreasing with $R$. Figure~\ref{fig:one} also shows that the azimuthal
speed of the gas in the ADAF (or equivalently its angular rotation
speed) is substantially sub-Keplerian ($\Omega \simeq 0.2 \Omega_K$ in
the present models).

\subsection{Spectral models}

We compute the X-ray spectra using an extended version of the Raymond
\& Smith (1977) code. The code computes the Bremsstrahlung,
recombination and two-photon continua, as well as the emission in
spectral lines.  The effects of collisional excitations (see,
e.g. Pradhan, Norcross \& Hummer 1981 and Pradhan 1985),
recombinations to excited levels of H-like and He-like ions (see,
e.g., Mewe, Schrijver, \& Sylwester 1980) and dielectronic
recombination satellite lines (see, e.g. Dubau et al. 1981 and
Bely-Daubau et al. 1982) are all included in the calculation.  The
ionization state of the elements is computed self-consistently within
the code itself.  Given the typical range of densities and infall
velocities in the ADAF (see Fig.~\ref{fig:one}), the ionization time
is much smaller than the infall time and the ionization state of the
elements is very close to equilibrium. Note that in computing the
expected luminosities, we do not take into account the partial
obscuration by the central WD of the inner shells of the accretion
flow. This should not strongly affect our predictions because the
outer regions of the flow contribute significantly to the total
luminosity in the radially extended ADAF models presented here. The
high temperature regions close to the WD surface would be most
strongly affected and the fluxes from these regions (which may be so
hot that they do not contribute much to the line emission) would be
reduced by a factor two at most.

Let $E_m(r,\theta)$ be the emissivity in a given line, and let $v_{\rm
los}$ be the local component of the velocity along the line of sight
to the observer. The emission profile of a line as measured by an
observer is then given by
\beq \Phi(v) = \int_0^{2\pi}d\phi\int_0^\pi
d\theta\sin\theta \int_{r_{\rm in}}^{r_{\rm out}}dr\,r^2
E_m(r,\theta)\;\Phi'(v;\, r,\theta,\phi) \;\;\;\;{\rm ergs} \;{\rm
s}^{-1}\; ({\rm km}\;{\rm s}^{-1})^{-1}\;,
\label{eq:prof}
\eeq
where 
\beq
\Phi'(v;\, r,\theta,\phi) =\frac{1}{\sqrt{\pi}\Delta}
\exp\left\{-\frac{[v-v_{\rm los}(r,\theta,\phi)]^2}{\Delta^2}\right\}\;,
\label{eq:proff}
\eeq and $\Delta=\sqrt{2k_{\rm B}T/m_a + v^2_{\rm turb}}$ for an atom
of mass $m_a$. For the turbulent velocity, we adopt $v_{\rm
turb}\simeq \alpha_{\rm ADAF} c_s$, where $c_s=(\gamma\mu k_B
T/m_p)^{1/2}$ is the fluid adiabatic sound speed and $\mu=0.5$ its
mean molecular weight. In the models presented below, we find that
line broadening is largely dominated by the flow bulk motion, with
only a small contribution from the turbulent motions (the contribution
from the thermal motions is negligible for the heavy element X-ray
lines discussed here; this also implies that turbulence is supersonic
for the heaviest species).

\section{Results}

\subsection{Continuum and Line Emission}

Whereas the emission of thermal X-ray lines is a robust prediction of
models of hot accretion flows, the detailed spectrum is not. It
depends on several parameters of the model, such as the radial extent
and the run of density in the hot flow. As we show below, this
property can be used, in turn, to constrain the structure of the hot
flow from detailed X-ray spectroscopic diagnostics.

The X-ray spectra corresponding to models I and II are shown in panels
(a) and (b) of Figure~2 . The spectra are characterized by emission
lines of Fe, O, Si and other heavy elements, superposed on the
Bremsstrahlung continuum.  Note that a solar abundance has been
assumed for the elements in the gas. Deviations from solar abundances
could result in the lines of specific elements being weaker or
stronger than in the models presented here. The shape of the continuum
and line emission contains information on the structure of the hot flow. As a
sum of the emission of shells of various densities $\rho$ and
temperatures $T$ (see Fig.~\ref{fig:one}), a careful fit to an
observational spectrum should provide valuable constraints on $\rho$
and $T$ in the hot flow. For instance, it is clear from Fig. 2 that
there is additional low-energy Bremsstrahlung continuum emission in
model II relative to model I, because of the additional emission from
the outer and cooler regions of the flow.

The relative importance of the various lines in the spectrum is also
indicative of the density, temperature and radial extent of the hot
flow. For the ADAF models, lines formed at temperatures $\la 10^8$~K
are absent from the emission spectrum if the radial extent of the flow
is $\la 10^4 R_s$ (Fig. 2, panel a), while they become clearly
apparent if the outer radius of the flow corresponds to a temperature
$\sim 10^7$ K (model shown in Fig. 2, panel b).  Similarly, in a
model where the accretion rate in the ADAF is not decreasing with
radius beyond $R_{\rm trans}=10^4 R_s$ (like in model II), but is
constant with radius, lines preferentially emitted at temperatures
$\la 10^8$~K would be stronger because of the higher densities in the
outermost regions of the flow. This illustrates that the relative
strengths of lines formed at various radii can be used to 
constrain the radial extent of the flow and its density
profile\footnote{ Note that this can be done even for unknown
elemental abundances, by considering lines from different ions of the
same element.}  (see Perna, Raymond \& Narayan 2000 for details).

This can be complemented by measurements of line profiles, which
strongly constrain the bulk motion of the hot gas in the flow. A
crucial advantage that DN offer over other classes of systems that
could harbor a hot flow is the (usually) relatively good knowledge of
the mass (and therefore the size) of the accreting WD and of the
inclination of the binary system. This provides an absolute scaling
for the width of the emission lines, provided the line-of-sight
velocities are known.  Note that, except for small values of the
inclination angle, broadening by the flow bulk motion is expected to
dominate over thermal broadening for determining the width of the
thermal X-ray lines in an ADAF.

Panels (a) and (b) of Figure 3 show the profiles of two strong lines,
from iron and oxygen atoms respectively, for the inclination angle
$i=40^o$ of SS Cyg. The predicted width of the lines is $\la
500$~km~s$^{-1}$.  Note that the oxygen line is narrower than the iron
line because it is produced in cooler regions of the ADAF, where it
rotates more slowly.  The width of these lines is a direct probe of
the sub-Keplerian rotation in the hot flow because the inclination of
the system (and to some extent the region of emission) is known. In
particular, if the rotation rate of the flow were Keplerian, lines
would be expected to be $\sim 5$ times broader. Since the
sub-Keplerian rotation is due to partial radial pressure support by
the hot gas, and this gas is hot because of energy advection, 
detection of these ``narrow'' lines would indicate the presence of
energy advection in the flow.

\subsection{Detectability with Current X-ray Instruments}

At a distance of $\simeq 166$~pc (Harrison et al. 1999), the typical
X-ray flux of SS Cyg ($0.5-10$~keV) is
$10^{-11}$~erg~cm$^{-2}$~s$^{-1}$ or more (Mukai \& Shiokawa 1993).
This allows a high quality spectrum to be taken in a few tens of ks of
observation with the powerful X-ray satellites {\it Chandra} and {\it
XMM-Newton}.  The two satellites have spectroscopic instruments which
approximately cover the soft and hard X-ray energy band corresponding
to the spectra shown in Fig. 2. If a radially extended ADAF is present
in SS Cyg, a spectrum taken by either of them should therefore yield
valuable constraints on the general structure of the hot flow (radial
extent, typical densities).

An important question is whether or not the width of the predicted
lines (see Fig. 3) can be resolved by current instruments. The
resolving power of the HETG on {\it Chandra} at 6.9~keV is 200
(corresponding to $1500$~km~s$^{-1}$), which is of order 3 times the
FWHM of the Fe XXVI line expected from the ADAF. This line cannot be
resolved, but it should be possible, from a high quality spectrum, to
rule out a line which is 5 times wider than predicted for the ADAF
model. It should therefore be possible to test the Sub-Keplerian
character of the hot flow with this instrument. Similarly, the
resolving power of the HETG around 1~keV is 1000 (corresponding to
$300$~km~s$^{-1}$), which is comparable to the width of the OVIII line
expected from the ADAF. If such a line is detected at high enough S/N,
it should be possible to tell with confidence if the hot emitting gas
rotates at a substantially sub-Keplerian rate or not.

The two EPIC cameras onboard {\it XMM-Newton} have a resolving power
of $\simeq 15$ at 1~kev and $\simeq 50$ at 6.4~keV, which is
insufficient for the line diagnostics envisioned here. However, the
RGS instrument (covering only the soft X-rays, from 0.35 to 2.5~keV)
has a resolving power of $\simeq 300-500$ which should be sufficient
to distinguish between a Keplerian and a substantially sub-Keplerian
rotation rate. A possible advantage of RGS over the {\it Chandra}
instruments is its larger sensitivity at low energies, allowing
spectroscopy with higher time resolution.

\section{Discussion}

There is already evidence in the literature for X-ray line emission
from DN in general and SS Cyg in particular. Mukai \& Shiokawa (1993)
find, from EXOSAT observations, that an emission line near 6.7~keV
originates from many DN, including SS Cyg. Yoshida et al. (1992) find
with the Ginga satellite that the X-ray emission of SS Cyg in
quiescence shows an emission line around 6.68~keV (see also Done \&
Osborne 1997). These observations are therefore encouraging in
suggesting the presence of a hot flow around the WD.  Their quality,
however, is insufficient to carry out the detailed spectroscopic
diagnostics described above.

These diagnostics will become possible with the new powerful X-ray
satellites {\it Chandra} \& {\it XMM-Newton}.  It should be possible,
from high quality X-ray spectra, to discriminate between the case of
accretion via an ADAF and the presence of a disk boundary layer (BL)
in a quiescent DN. To our knowledge, there is no detailed spectral
prediction for the emission originating from a compact BL (at low
accretion rates), but one expects emission from a much denser gas than
in the radially extended ADAF, possibly with temperatures ranging from
the keV range all the way down to the WD atmospheric temperature at
$15,000$~K or so.  In first approximation, the dissipation of the
rotational kinetic energy of the gas in the boundary layer leads to
temperatures $T$ such that $3 k_B T \sim 1/2 m_p (V_K^2-V_*^2)$, where
$V_K$ is the Keplerian speed in the disk and $V_*$ is the speed at the
stellar surface. If $T$ drops toward the WD surface, lines from cooler
regions should have smaller rotational widths. For example, the model
of Mahasena \& Osaki (1999) does not treat the azimuthal velocity
explicitly, but it predicts that most of the energy liberated in the
boundary layer is transported to a dense region close to the WD
surface and radiated away at intermediate temperatures.  Because the
radiating gas is the part of the BL closest to the stellar surface,
its angular speed should be close to that of WD, and the line widths
should be even smaller than in the ADAF case. However, broadening by
turbulent motions and the fact that detailed and more realistic
calculations show that energy is radiated radially outward and
advected radially inward in BLs (e.g. Narayan \& Popham 1993; Popham
\& Narayan 1995; Godon, Regev \& Shaviv 1995; Regev \& Bertout 1995)
could modify this simple picture quite substantially.

A possibly important difference between an ADAF model and simple BL
models (in which all the kinetic energy of the accreting gas is
radiated as one-temperature Bremsstrahlung emission and the role of
advection is neglected; e.g. Patterson \& Raymond 1985a) is in the
ionization state of the gas, as manifested in the emission line
equivalent widths.  The fundamental difference is that only a small
fraction of the accretion luminosity in the ADAF model emerges as
fairly energetic X-rays, while most of the luminosity of such a simple
boundary layer does.  Thus photoionization may be significant and the
ionization parameter higher in the simple BL model.  Elements such as
oxygen will be more highly ionized, with O~VII and O~VIII replaced by
O~IX.  Thus the O~VII $\lambda$22 and O~VIII $\lambda$19 lines should
be weaker in the simple BL model than in an ADAF model.  In that
respect, the models presented in this study (specific to SS Cyg) have
general properties which are representative of what is expected from
ADAFs possibly present in other quiescent DN.  More specific BL model
predictions for emission line fluxes and profiles are badly needed.

The spectral models presented in this study are for specific values of
the ADAF model parameters (such as $\alpha_{\rm ADAF}=0.2$ and $\dot
M= 7 \times 10^{-3} \dot M_{\rm Edd}$~$\simeq 1.5 \times
10^{-10}~M_\odot$~yr$^{-1}$). We find, however, that the spectral
predictions would be only marginally affected, or affected in trivial
ways, if these parameters were changed, so that we felt it was
unnecessary to explore here the parameter space of the ADAF
models. For example, line profiles are sensitive to the value of
$\alpha_{\rm ADAF}$ only for small values of the
inclination\footnote{It is only the radial component of the flow
velocity that depends on $\alpha$, but not the rotational one. At
small inclinations, turbulent motions become the main source of line
broadening, together with thermal motions.}, while line ratios are
completely insensitive to changes in $\alpha_{\rm ADAF}$\footnote{The
run of density is independent of $\alpha_{\rm ADAF}$ (only the overall
scaling depends on it), and the run of temperature does not directly
depend on the value of $\alpha_{\rm ADAF}$ (it depends only on the
value of $\beta$; see, e.g., Menou 2000).}. Changes in the values of
$\dot M$ do affect the overall spectrum, by rescaling the overall
continuum and line emission, but without relative changes between the
lines.  The rate of rotation, on the other hand, depends rather
sensitively on the adiabatic index $\gamma$ of the fluid. Note,
however, that even if extreme values are assumed for this variable in
an ADAF, the flow still remains sub-Keplerian ($\Omega \la 0.5
\Omega_{\rm K}$ when $\gamma \rightarrow 4/3$; Narayan \& Yi 1995) and
our conclusions are mainly unchanged.

A possible ingredient of the X-ray spectrum of a quiescent DN that has
been omitted from our calculations is the iron $K_\alpha$ fluorescence
line at $6.4$~keV from the X-ray irradiated disk. This line can appear
on top of the Bremsstrahlung continuum emission, depending on the
radial extent of the truncated disk. The width of the line, related to
the Keplerian rotation rate in the inner regions of the disk and the
system inclination is $\sim$ a few $10$~eV at most (Menou 2000).  This
is still narrow enough that the line should be easily separable, in a
good quality spectrum, from the other iron lines expected at $6.7$ and
$6.9$~keV (from the hot gas).

Finally, it is important to notice that there are possible
complications in interpreting narrow X-ray emission lines in terms of
energy advection in a hot flow. One possible source of confusion comes
from lines emitted in the close vicinity of the WD (for instance in a
disk BL scenario), which is also likely to be rotating at a
sub-Keplerian rate (see Sion 1999 for existing constraints on the
rotation rate of WDs in DN). The sub-Keplerian rotation of the
emitting gas can therefore be associated to the property of energy
advection of the hot flow only if it can be shown that it is not
directly related to the slow rotation of the WD. This may require
clear evidence that the hot flow is indeed radially extended.  We also
note that Medvedev \& Narayan (2000) recently presented solutions for
accretion onto a compact object in which the flow can be radially
extended and have a substantially sub-Keplerian rotation rate, without
advecting any energy.\footnote{The rotation rate is directly related
to that of the central object in these solutions.} If such a hot flow
were present around WDs in quiescent DN, it could also be at the
origin of narrow X-ray emission lines. The interpretation of narrow
lines in terms of energy advection will therefore require some
knowledge of other properties of the flow, such as its density
profile, which differs in the ADAF solutions and the solutions of
Medvedev \& Narayan (2000). The knowledge of the WD rotation rate in
some DN and the possibility of actually resolving some of the low
energy X-ray lines predicted for ADAFs (\S4.2) should provide critical
observational tests of these two accretion scenarios.

\section{Conclusion}

We have constructed detailed spectral models of an ADAF around a WD,
with model parameters appropriate for the Dwarf Nova SS Cygni in
quiescence. The ADAF is a source of X-ray Bremsstrahlung continuum
emission, which is characteristic of its density, temperature and
radial extent.

Strong X-ray lines are also emitted by highly ionized atoms of Fe, O,
Si and other heavy elements in the flow. The pattern of emission lines
is a robust probe of the density, temperature and elemental abundances
in the hot flow, which can therefore be constrained from high quality
X-ray spectra.

In addition, the width of the X-ray emission lines can be directly
related to the bulk motion of the gas in the ADAF by using the {\it a
priori} knowledge of the mass of the WD and the inclination to the
line of sight of the system.  Current X-ray instruments have the
capabilities to distinguish between lines emitted from a flow rotating
at or substantially below the Keplerian rate. The property of energy
advection of the flow, which is at the origin of the sub-Keplerian
rotation in an ADAF, can therefore in principle be tested accurately
by the observations.

\section*{Acknowledgments}
We are grateful to Jonathan McDowell for many enlightening discussions
on the observational capabilities of {\it Chandra}. Support for this
work was provided by NASA through Chandra Postdoctoral Fellowship
grant number PF9-10006 awarded by the Chandra X-ray Center, which is
operated by the Smithsonian Astrophysical Observatory for NASA under
contract NAS8-39073. Partial support was provided by NASA Grant
NAG-528 to the Smithsonian Astrophysical Observatory.

\clearpage

\clearpage
\begin{figure}
\plotone{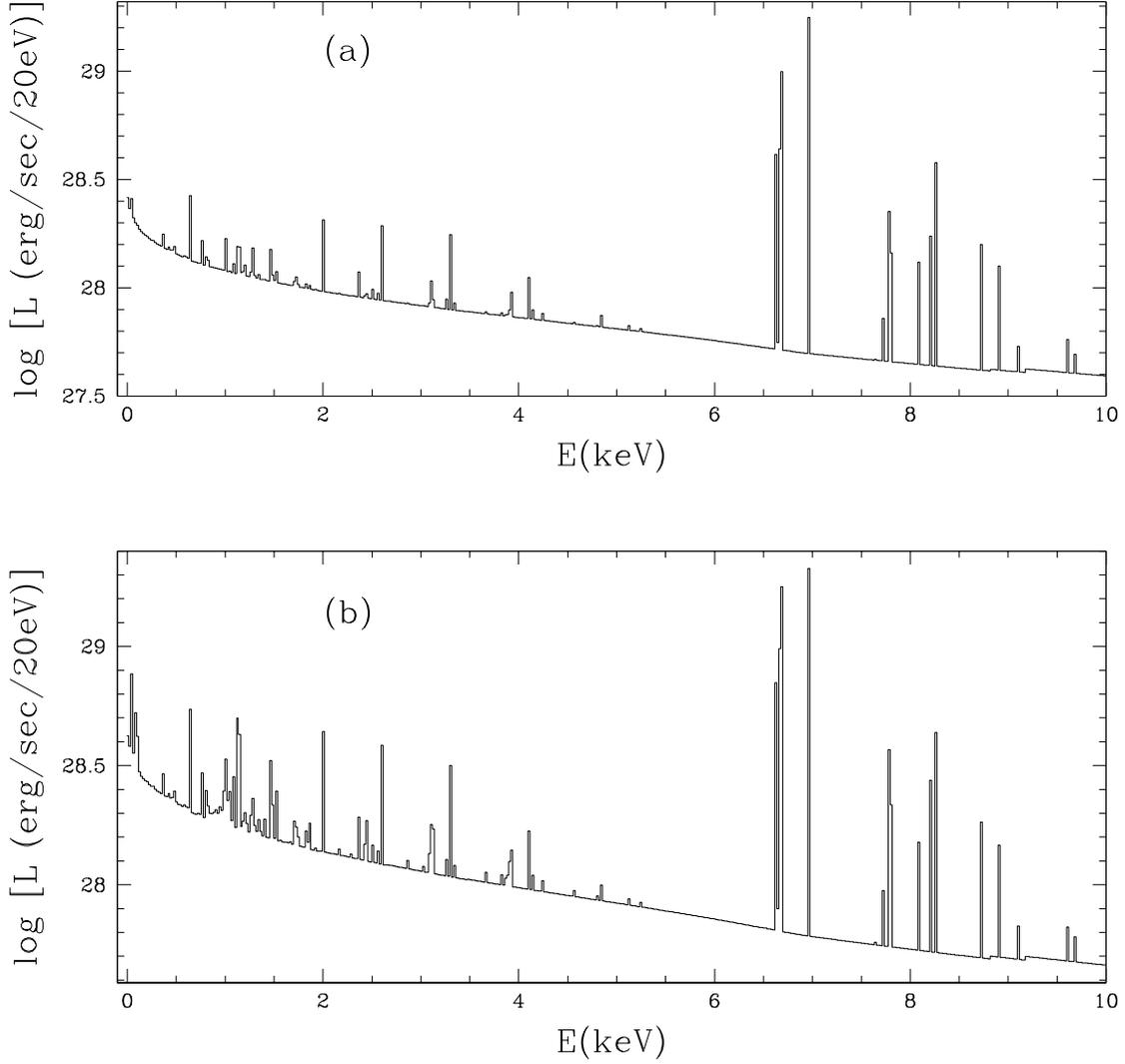}
\caption{The predicted X-ray continuum and line emission for the ADAF
models I (a) and II (b). In model II, the ADAF extends further out,
where the gas is cooler and emits additional low-energy lines and
Bremsstrahlung continuum. The various lines in (a) and (b) correspond
to emission from atoms of Fe, O, Si and other heavy elements.
\label{fig:two}}
\end{figure}

\clearpage
\begin{figure}
\plotone{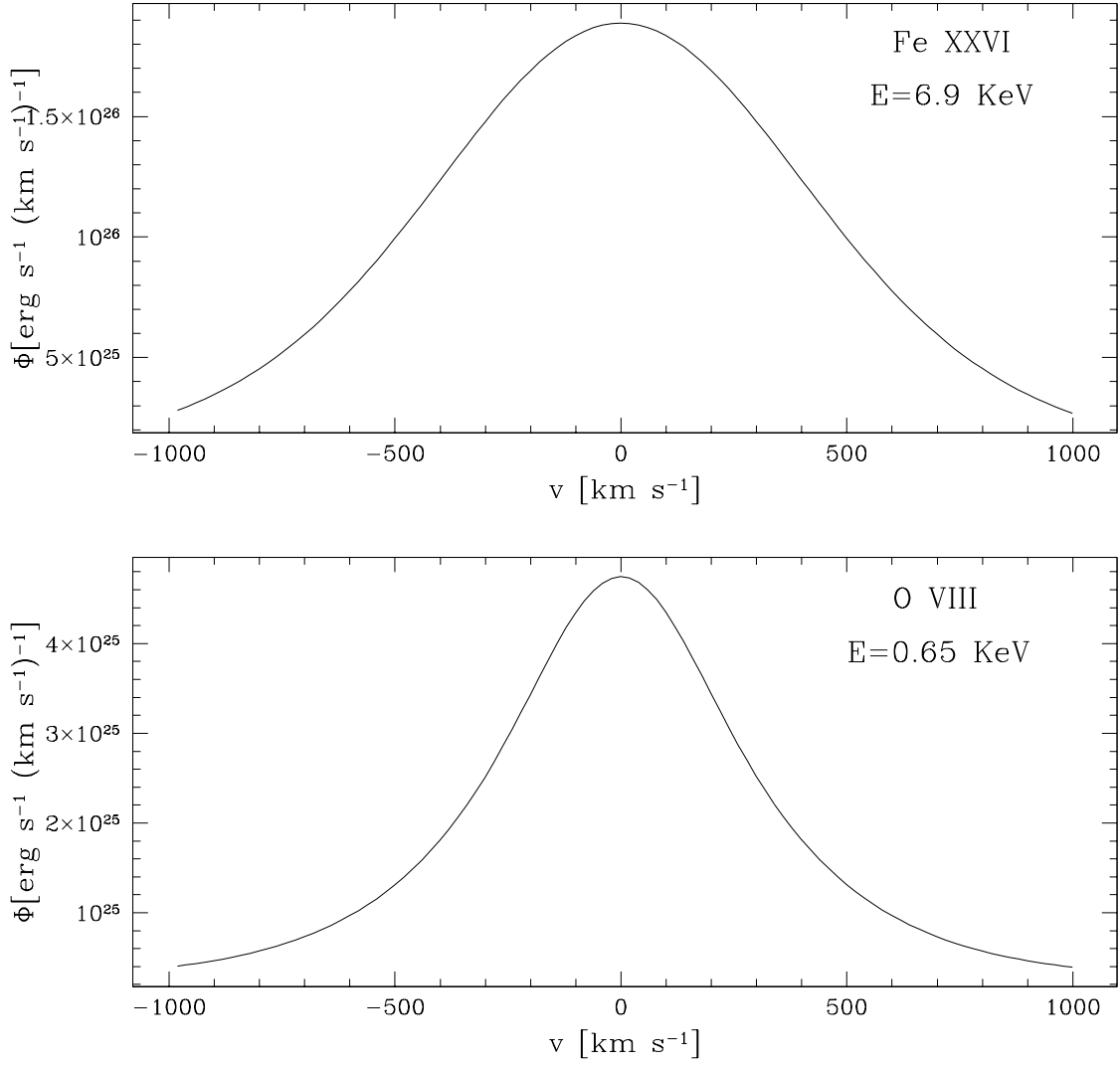}
\caption{The predicted profiles for the strong iron line at 6.9~keV
and the strong oxygen line at 0.65~keV in model II.  The oxygen line
is narrower than the iron line because it originates from cooler
regions of the flow, where it rotates more slowly. 
\label{fig:two}}
\end{figure}

\end{document}